\documentclass[a4paper]{jpconf}
\usepackage{graphicx,epsfig}
\begin{document}
\title{New limits on $2\beta$ processes in $^{106}$Cd}

\author{
V.I.~Tretyak$^{1,2}$, 
P.~Belli$^{3}$, 
R.~Bernabei$^{3,4}$, 
V.B.~Brudanin$^{5}$,
F.~Cappella$^{6}$,  
V.~Caracciolo$^{6}$, 
R.~Cerulli$^{6}$, 
D.M.~Chernyak$^{1}$,
F.A.~Danevich$^{1}$, 
S.~d'Angelo$^{3,4,*}$, 
A.~Di~Marco$^{4}$, 
A.~Incicchitti$^{2,7}$, 
M.~Laubenstein$^{6}$,
V.M.~Mokina$^{1}$,  
D.V.~Poda$^{1,8}$, 
O.G.~Polischuk$^{1,2}$, 
I.A.~Tupitsyna$^{9}$
}

\address{$^1$ Institute for Nuclear Research, MSP 03680 Kyiv, Ukraine}
\address{$^2$ INFN, sezione di Roma, I-00185 Rome, Italy}
\address{$^3$ INFN, sezione di Roma ``Tor Vergata'', I-00133 Rome, Italy}
\address{$^4$ Dipartimento di Fisica, Universit\`{a} di Roma ``Tor Vergata'', I-00133 Rome, Italy}
\address{$^5$ Joint Institute for Nuclear Research, 141980 Dubna, Russia}
\address{$^6$ INFN, Laboratori Nazionali del Gran Sasso, I-67100 Assergi (AQ), Italy}
\address{$^7$ Dipartimento di Fisica, Universit\`{a} di Roma ``La Sapienza'', I-00185 Rome, Italy}
\address{$^8$ Centre de Sciences Nucl\'{e}aires et de Sciences de la Mati\`{e}re, 91405 Orsay, France}
\address{$^9$ Institute of Scintillation Materials, 61001 Kharkiv, Ukraine}
\address{$^*$ Deceased}

\ead{tretyak@kinr.kiev.ua}

\begin{abstract}
A radiopure cadmium tungstate crystal scintillator, enriched in
$^{106}$Cd to 66\%, with mass of 216 g ($^{106}$CdWO$_4$) 
was used in coincidence with four ultra-low background HPGe detectors 
contained in a single cryostat
to search for double beta ($2\beta$) decay processes in $^{106}$Cd.
New improved half-life limits on the $2\beta$
processes in $^{106}$Cd have been set on the level of
$10^{20}- 10^{21}$ yr after 13085 h of data taking deep underground (3600 m w.e.)
at the Gran Sasso National Laboratories of INFN (Italy). In particular,
the limit on the two neutrino electron capture with
positron emission, $T_{1/2}^{\varepsilon\beta^+ 2\nu}\geq 1.1\times
10^{21}$~yr, has reached the region of theoretical predictions.
The resonant
neutrinoless double electron captures to the 2718, 2741
and 2748 keV excited states of $^{106}$Pd are restricted on the
level of $T_{1/2}^{2\varepsilon 0\nu} \geq (8.5\times10^{20}-1.4\times10^{21}$) yr.
\end{abstract}

\section{Introduction}

While we already know from experiments on neutrino oscillations that $\nu$'s have non-zero masses,
their absolute values are unknown because these investigations are sensitive only to differences 
in $\nu$ mass squares \cite{Pat15}. Experiments on neutrinoless ($0\nu$) double beta ($2\beta$) decay
of atomic nuclei $(A,Z)$ $\to$ $(A,Z\pm2) + 2e^\mp$
are considered to-date as the only reliable way to find the mass absolute 
scale and to study the neutrino properties (are they Majorana, $\nu=\bar{\nu}$, or Dirac, 
$\nu\neq\bar{\nu}$, particles). This process is related also with other effects
beyond the Standard Model (SM), like possible existence of right-handed currents in weak interaction,
Majorons, etc. In spite of searches for $2\beta0\nu$ decay during near 70 years,
it is still not surely observed, with half-life sensitivities of $\simeq 10^{25}$ yr for 
$(A,Z)$ $\to$ $(A,Z+2)$ and $\simeq 10^{21}-10^{22}$ yr for $(A,Z)$ $\to$ $(A,Z-2)$ processes 
reached in the best experiments. 
Two neutrino ($2\nu$) mode of $2\beta$ decay (process allowed in the SM)
was already observed in more than 10 nuclides
with $T_{1/2} \simeq 10^{18}-10^{24}$ yr; see the last reviews 
\cite{Pas15,Sar15,Bil15,Cre14,Maa13,Saa13,Sch13,Giu12,Ver12,Fae12,Vog12,Ell12,Gom12} 
and refs. therein.

$^{106}$Cd is one of the best candidates to search for processes $(A,Z)$ $\to$ $(A,Z-2)$:
double positron emission ($2\beta^+$),
electron capture with positron emission ($\varepsilon\beta^+$) and
two electron capture ($2\varepsilon$)
because of its high decay energy ($Q_{2\beta}=2775.39(10)$ keV \cite{Wan12})
and comparatively high natural abundance ($\delta=1.25(6)\%$ \cite{Ber11}).
Investigations of $2\beta^+ / \varepsilon\beta^+ / 2\varepsilon$ processes could clarify
question about possible contribution of right-handed admixtures in the weak
interaction to $2\beta0\nu$ decay probability \cite{Hir94}.
$^{106}$Cd nucleus is interesting also because of possible resonant $2\varepsilon 0\nu$ captures
to excited levels of the daughter nucleus $^{106}$Pd which could be enhanced by few
orders of magnitude because of proximity of the released energy to energy of one or more
of the excited levels \cite{Maa13,Kri11}.

In recent searches for $2\beta$ decay of $^{106}$Cd, 32 planar HPGe detectors and 16
thin $^{106}$Cd metallic foils between them were used in the TGV-2 experiment \cite{Bri15}, and array
of CdZnTe room temperature semiconductors was used in the COBRA studies \cite{Ebe13}. 
At the first stage of our investigations \cite{Bel12}, $^{106}$CdWO$_4$ crystal enriched
in $^{106}$Cd to 66\% with mass of 216 g was used as a scintillating detector. 
At the second stage, described here, it is operated in low background set-up together with four
HPGe detectors enhancing sensitivity to some $2\beta$ processes with emission of 
$\gamma$ quanta. We report here preliminary results of the experiment.

\section{Experimental set-up and measurements}

The $^{106}$CdWO$_4$ scintillator ($\oslash27\times50$ mm, mass 216 g) was grown from
deeply purified Cd (66\% of $^{106}$Cd) by the Low-Thermal-Gradient Czochralski method
\cite{Bel10}.
It was optically connected to a low-background photomultiplier tube (PMT, Hamamatsu
R6233MOD) through a radiopure PbWO$_4$ crystal light-guide ($\oslash40\times83$ mm)
produced from deeply purified archaeological lead that allowed to suppress radioactivity 
from PMT. The detector was installed in an ultra-low background GeMulti HPGe 
spectrometer at the Gran Sasso underground laboratory (LNGS) of the INFN (Italy)
at the depth 3600 m w.e.
Four HPGe detectors (with volumes approximately 225 cm$^3$ each) were mounted 
in one cryostat with a well in the centre. 
An event-by-event data acquisition system stored the time of arrival of the events in 
the $^{106}$CdWO$_4$ and HPGe detectors, and the pulse shape of $^{106}$CdWO$_4$ 
scintillation signals. The $^{106}$CdWO$_4$ and HPGe detectors were calibrated
with $^{22}$Na, $^{60}$Co, $^{137}$Cs and $^{228}$Th.
The energy resolution of the $^{106}$CdWO$_4$ detector can be described by the 
function: FWHM = $\sqrt{21.7 \times E_\gamma}$, with FWHM and $E_\gamma$ in keV. 
The energy resolution of the HPGe spectrometer is $\simeq 2.0$ keV for the 1332 keV 
$\gamma$ quanta of $^{60}$Co. More details are given in \cite{Bel16}.

The data were accumulated during 13085 h. 
The pulse-shape discrimination based on the mean time of the scintillation signal \cite{Bar06}
was applied to discriminate events caused by $\gamma$ and $\beta$ particles from those
induced by $\alpha$'s.
Fig.~1 (left) shows $^{106}$CdWO$_4$ energy spectra:
in anti\-co\-in\-cidence with HPGe detectors; 
in coincidence when energy release in at least one of the HPGe detectors is $>200$ keV; 
and in coincidence when $E$(HPGe) = 511 keV ($\pm3\sigma$, 
where $\sigma$ is the energy resolution of the HPGe detectors at 511 keV). 

\begin{figure}[htb]
\mbox{\epsfig{figure=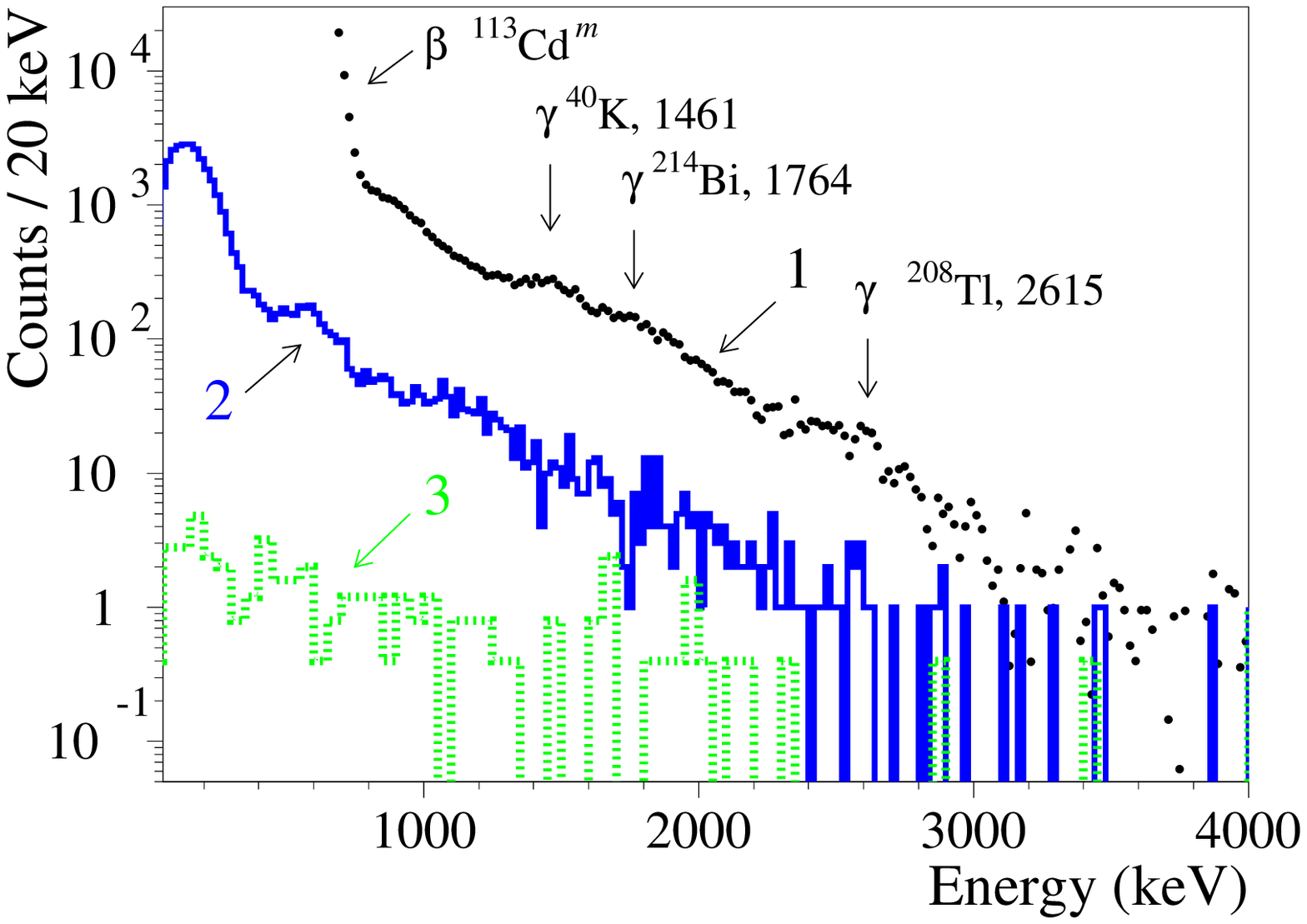,height=5.7cm}}~\mbox{\epsfig{figure=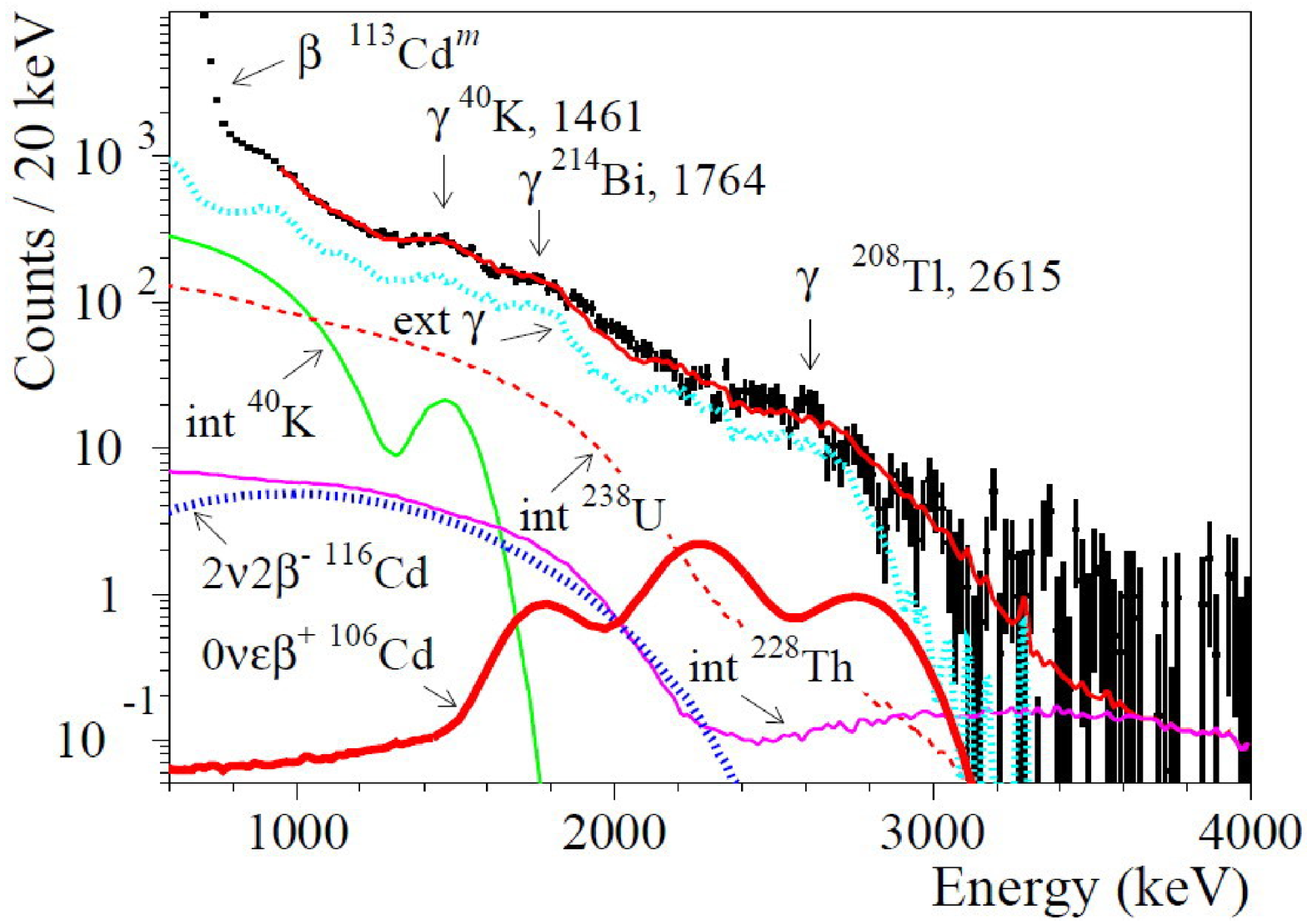,height=5.7cm}}
\caption{
Left: $^{106}$CdWO$_4$ energy spectra collected during 13085 h:
1 -- in anticoincidence with HPGe detectors;
2 -- in coincidence when energy release in at least one of the HPGe is $>200$ keV;
3 -- in coincidence when $E$(HPGe) = $511 (\pm 3\sigma)$ keV.
Right: Fit of the anticoincidence spectrum by background model (red 
continuous line), and its main components. The excluded distribution of the 
$\varepsilon\beta^+0\nu$ decay of $^{106}$Cd to the ground state of $^{106}$Pd with 
$T_{1/2} = 1.5\times10^{21}$ yr is shown too.
}
\end{figure}


\section{Results and discussion}

Contributions of possible radioactive sources to the collected spectra were simulated
with the EGS4 code \cite{Nel85}. The list includes, in particular,
radioactive contaminations of the $^{106}$CdWO$_4$ crystal scintillator \cite{Bel12}, 
external $\gamma$ quanta from the PMT and materials of the set-up, 
and also $2\beta2\nu$ decay of $^{116}$Cd present in the $^{106}$CdWO$_4$ crystal on the 
level of 1.5\% \cite{Bel10}. Fit of the $^{106}$CdWO$_4$ anticoincidence spectrum by 
the background model, and its main components are shown in Fig.~1 (right).

Responce of the $^{106}$CdWO$_4$ scintillator to different modes of $2\beta$ decay
of $^{106}$Cd to the ground state and excited levels of $^{106}$Pd were also
simulated with the EGS4; initial kinematics of particles emitted in decay and
deexcitation of the daughter nucleus was given by the DECAY0 event generator 
\cite{Pon00}. In general, we did not find any peculiarities in the data 
accumulated with the $^{106}$CdWO$_4$ and HPGe detectors that could be
ascribed to the $2\beta$ processes in $^{106}$Cd.
Thus we give only the half-life limits according to a formula:
$\lim T_{1/2} = \ln 2 \cdot N \cdot \eta \cdot t / \lim S$,
where $N$ is the number of $^{106}$Cd nuclei in the $^{106}$CdWO$_4$ crystal 
($N = 2.42 \times 10^{23}$), 
$\eta$ is the detection efficiency, $t$ is the time of measurements, 
and $\lim S$ is the number of events of the effect searched for, which
can be excluded at a given confidence level (C.L.).

We have analyzed different data to estimate limits on the $2\beta$
processes in $^{106}$Cd. For instance, to derive the value of
$\lim S$ for the $\varepsilon\beta^+0\nu$ decay of $^{106}$Cd to
the ground state of $^{106}$Pd, the  $^{106}$CdWO$_4$
anticoincidence spectrum was fitted by the model built from the
components of the background and the effect searched for. The best
fit, achieved in the energy interval $1000-3200$ keV, gives the
area of the effect $S=27\pm 49$ counts, thus providing no evidence
for the effect. In accordance with the Feldman-Cousins procedure
\cite{Fel98}, this corresponds to $\lim S=107$ counts at 90\% C.L.
Taking into account the detection efficiency within the interval
given by the Monte Carlo simulation (69.3\%) and the 95.5\%
efficiency of the pulse-shape discrimination to select $\gamma$
and $\beta$ events, we got the half-life limit: $T_{1/2}\geq
1.5\times 10^{21}$ yr. The excluded distribution of the
$\varepsilon\beta^+0\nu$ decay of $^{106}$Cd to the ground state
of $^{106}$Pd is shown in Fig.~1 (right). 

The counting rate of the $^{106}$CdWO$_4$ detector is
substantially suppressed in coincidence with the energy 511 keV in
the HPGe detectors. The coincidence energy spectrum of the
$^{106}$CdWO$_4$ detector is presented in Fig.~2 (left). There
are only 115 events in the energy interval $0.05-4$ MeV, while the
simulated background model (built by using the parameters of the
anticoincidence spectrum fit) contains 108 counts. We have
estimated values of $\lim S$ for the $2\beta$ processes in
$^{106}$Cd in different energy intervals. Some of the excluded 
distributions are presented in Fig.~2 (left); corresponding 
$T_{1/2}$ limits are given in Table~1.
In particular, the half-life limit on the $\varepsilon\beta^+2\nu$ decay 
is equal $T_{1/2}\geq 1.1\times10^{21}$~yr.
This value is close to theoretical predictions of \cite{Sto03} where 
$T_{1/2} = (1.4-1.6)\times10^{21}$ yr was calculated.

Using the relation between the effective nuclear matrix element (NME)
for $\varepsilon\beta^+2\nu$ decay:
$(T_{1/2}^{\varepsilon\beta^+2\nu})^{-1} = G^{\varepsilon\beta^+2\nu} \cdot
|M^{\varepsilon\beta^+2\nu}|^2$,
and recent calculations of phase space factor 
$G^{\varepsilon\beta^+2\nu} = (702-741)\times10^{-24}$ yr \cite{Kot13,Mir15},
one can obtain a limit on NME for $\varepsilon\beta^+2\nu$ decay of $^{106}$Cd to
the ground state of $^{106}$Pd as:
$M^{\varepsilon\beta^+2\nu} < 1.1.$ 

We also used the data accumulated by the HPGe detectors to
estimate limits on the $2\beta$ processes in $^{106}$Cd. For
instance, in neutrinoless $2\varepsilon$ capture we assume that
the energy excess is taken away by bremsstrahlung $\gamma$ quanta
with energy $E_\gamma = Q_{2\beta} - E_{b1} - E_{b2} - E_{exc}$,
where $E_{bi}$ is the binding energy of $i$-th captured electron
on the atomic shell, and $E_{exc}$ is the energy of the populated
(g.s. or excited) level of $^{106}$Pd. In case of transition to an
excited level, in addition to the initial $\gamma$ quantum, other
$\gamma$'s will be emitted in the nuclear deexcitation process.
For example, to derive a limit on the $2K0\nu$ capture in
$^{106}$Cd to the ground state of $^{106}$Pd the energy spectrum
accumulated with the HPGe detectors was fitted in the energy
interval $2700-2754$ keV by a simple function (first degree
polynomial function to describe background plus Gaussian peak at
the energy 2726.7 keV with the energy resolution FWHM$~=4.4$ keV
to describe the effect searched for). The fit gives an area of the
peak $6.2\pm3.2$ counts, with no evidence for the effect.
According to \cite{Fel98} we took 11.4 events which can be
excluded with 90\% C.L. Taking into account the detection
efficiency for $\gamma$ quanta with energy 2726.7 keV in the
experimental conditions (1.89\%), we have set the following limit
for the $2K0\nu$ capture of $^{106}$Cd to the ground state of
$^{106}$Pd: $T_{1/2}\geq 4.2\times10^{20}$ yr. 
The excluded peaks are shown in Fig.~2 (right).

\begin{figure}[htb]
\mbox{\epsfig{figure=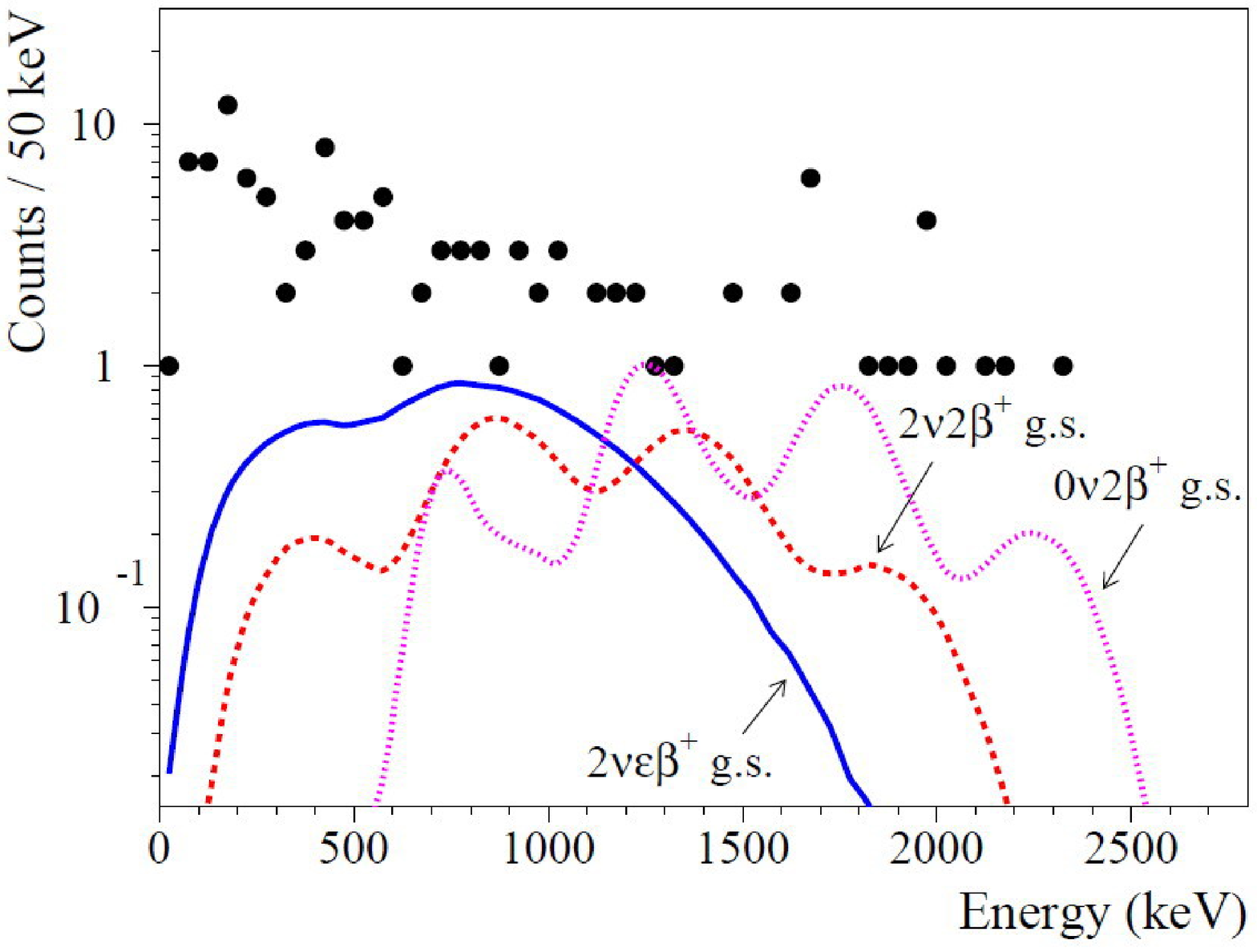,height=5.9cm}}~\mbox{\epsfig{figure=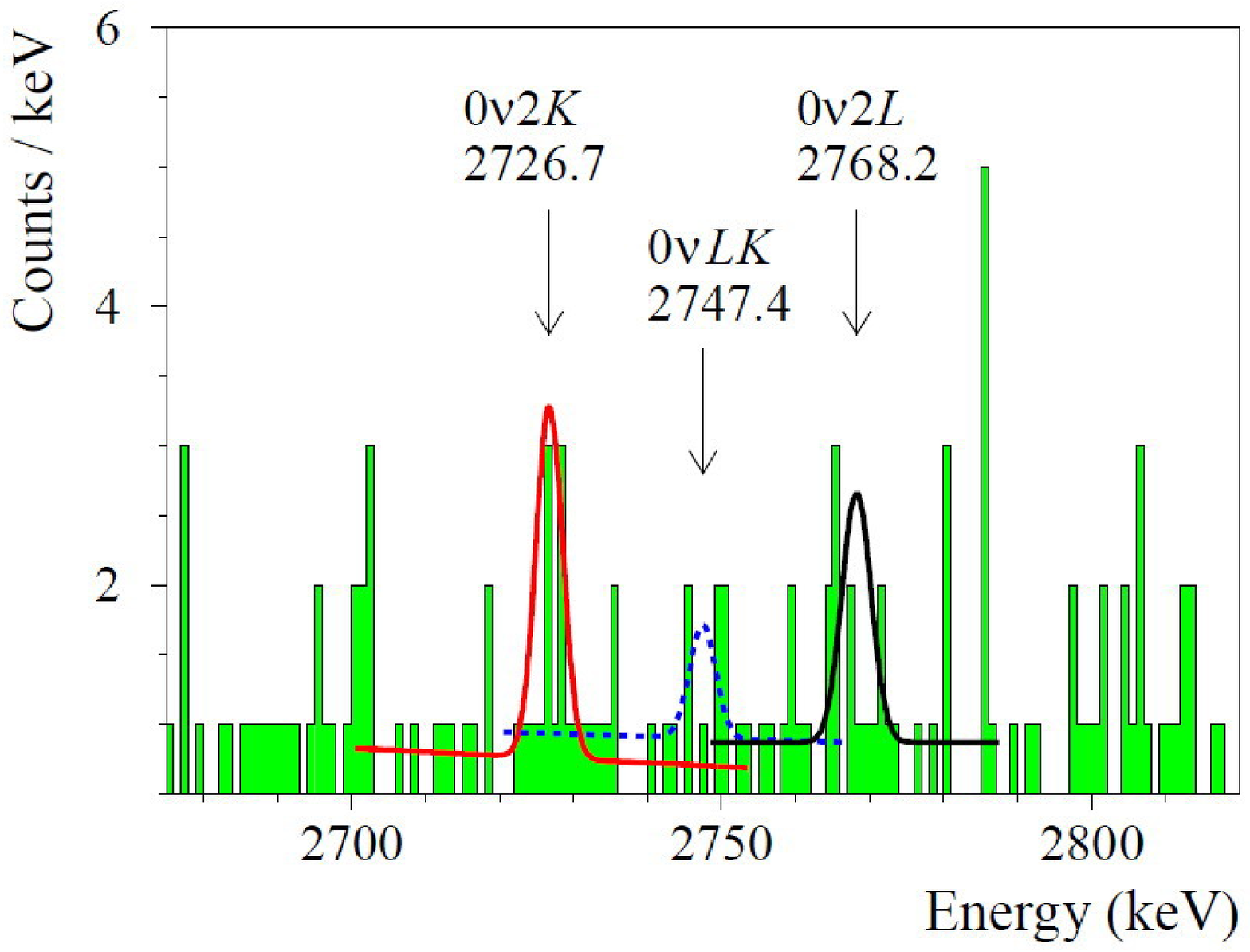,height=6.0cm}}
\caption{
Left: Energy spectrum of the $^{106}$CdWO$_4$ detector during 13085 h in coincidence 
with 511 keV annihilation $\gamma$ quanta in at least one of the HPGe detectors 
(filled circles). The excluded distributions of different $2\beta$ processes in 
$^{106}$Cd are shown by different lines.
Right: Part of the energy spectrum accumulated by the HPGe detectors.
Excluded peaks expected in the $2K0\nu$, $KL0\nu$ and $2L0\nu$ captures in $^{106}$Cd 
to the ground state of $^{106}$Pd are shown.
}
\end{figure}

Some of the obtained half-life limits on different $2\beta$ processes in $^{106}$Cd 
are given in Table 1, where results of the most sensitive previous experiments 
are also given for comparison.

\nopagebreak
\begin{table}[htb]
\caption{$T_{1/2}$ limits on 2$\beta$ processes in $^{106}$Cd
(AC -- anticoincidence with HPGe; CC -- coincidence with the given energy in HPGe; 
HPGe -- using data of only HPGe detectors).}
\begin{center}
\begin{tabular}{lll}
\hline
Decay and                               & \multicolumn{2}{c}{$T_{1/2}$ limit (yr) at 90\% C.L.}                        \\
$^{106}$Pd level (keV)                  & Present work (data)                   & Best previous limit                  \\
\hline
$2\beta^+0\nu$,           g.s.          & $\geq 3.0\times10^{21}$ (CC 511 keV)  & $\geq 1.2\times10^{21}$ \cite{Bel12} \\
$2\beta^+2\nu$,           g.s.          & $\geq 2.3\times10^{21}$ (CC 511 keV)  & $\geq 4.3\times10^{20}$ \cite{Bel12} \\
$\varepsilon\beta^+0\nu$, g.s.          & $\geq 1.5\times10^{21}$ (AC)          & $\geq 2.2\times10^{21}$ \cite{Bel12} \\
$\varepsilon\beta^+2\nu$, g.s.          & $\geq 1.1\times10^{21}$ (CC 511 keV)  & $\geq 4.1\times10^{20}$ \cite{Bel99} \\
$\varepsilon\beta^+2\nu$, $0^{+}$ 1134  & $\geq 1.1\times10^{21}$ (CC 622 keV)  & $\geq 3.7\times10^{20}$ \cite{Bel12} \\
$2K0\nu$,                 g.s.          & $\geq 4.2\times10^{20}$ (HPGe)        & $\geq 1.0\times10^{21}$ \cite{Bel12} \\
$2\varepsilon2\nu$,       $0^{+}$ 1134  & $\geq 1.0\times10^{21}$ (CC 622 keV)  & $\geq 1.7\times10^{20}$ \cite{Bel12} \\
Res. $2K0\nu$,            2718          & $\geq 1.1\times10^{21}$ (CC 1160 keV) & $\geq 4.3\times10^{20}$ \cite{Bel12} \\
Res. $KL_{1}0\nu$,        $4^+$ 2741    & $\geq 8.5\times10^{20}$ (HPGe)        & $\geq 9.5\times10^{20}$ \cite{Bel12} \\
Res. $KL_{3}0\nu$,        $2,3^-$ 2748  & $\geq 1.4\times10^{21}$ (CC 2236 keV) & $\geq 4.3\times10^{20}$ \cite{Bel12} \\
\hline
\end{tabular}
\label{2b-results}
\end{center}
\end{table}

\section{Conclusions}

An experiment to search for $2\beta$ decay of $^{106}$Cd with enriched 
$^{106}$CdWO$_4$ crystal scintillator with mass of 216 g in coincidence with four HPGe detectors
has been completed after 13085 h of data taking. New improved limits on 
$2\beta^+ / \varepsilon\beta^+ / 2\varepsilon$ processes in $^{106}$Cd were set on the level of 
$T_{1/2} > 10^{20}-10^{21}$ yr. The half-life limit on $\varepsilon\beta^+ 2\nu$ decay
$T_{1/2} > 1.1\times10^{21}$ yr reached the region of some theoretical predictions. 
Advancement of the experiment in the version when $^{106}$CdWO$_4$ scintilaltor is 
operating in coincidence with two large volume radiopure CdWO$_4$ scintillation detectors
in close (almost $4\pi$) geometry is in progress.

\section*{Acknowledgment}

The authors from the Institute for Nuclear Research (Kyiv, Ukraine) were supported in part 
by the project CO-1-2/2015 of the Program of collaboration with the Joint Institute 
for Nuclear Research (Dubna, Russia) ``Promising basic research in High Energy and 
Nuclear Physics'' for 2014-2015 of the National Academy of Sciences of Ukraine.

\end{document}